\documentstyle[12pt] {article}

\renewcommand{\theequation}{\arabic{section}.\arabic{equation}}
\renewcommand{\thefootnote}{\fnsymbol{footnote}}
\begin{document}
\baselineskip 24pt
%
\def\ev{\equiv}
\def\bra{\langle}
\def\ket{\rangle}
\def\bbra{\langle\!\langle}
\def\kket{\rangle\!\rangle}
\def\to {\vert}
\def\con{\ , \qquad}
\def\scon{\ ; \qquad}
\def\p  {\ .       }
\def\ov{\overline}
\def\wt{\widetilde}
\def\wh{\widehat}
\def\maru{\mathaccent "7017}
%
\hfill
\begin{minipage}[t]{3cm} 
DPNU-98-24 \\ June 1998 
\end{minipage}
%

\noindent
\hspace*{\fill}
\vspace*{1.5cm}
 \begin{center}
  {\large\bf On the Approximation in the Hermitian Treatment of 
Dyson Boson Expansion Theory
\renewcommand{\thefootnote}{\fnsymbol{footnote}}
\footnote [2] { submitted to Prog. Theor. Phys. }
}
\vspace{2cm}

  Atsushi K{\footnotesize AJIYAMA},
  Kimikazu T{\footnotesize ANIGUCHI}${}^{*}$
  and Yoshinao M{\footnotesize IYANISHI}${}^{**}$
  \\
  \vspace{1cm}

  {\it Department of Medical Information Science,
   \\
  Suzuka University of Medical Science, Mie 510-0293  }
   \\
  {\it${}^{*}$  Department of Medical Electronics,
   \\
  Suzuka University of Medical Science, Mie 510-0293} 
   \\
  {\it${}^{**}$ Graduate School of Science, Nagoya University, 
   Nagoya 464-8602}
 \end{center}
\vspace{3cm}
{\bf Abstract:}
We discuss about the Hermitian treatment of Dyson-type boson 
expansion theory.
We show that the basic assumption of the conventional treatment
does not hold in general and the method is only approximately
valid. 
The exception is the case where the multi-phonon states 
are mutually orthogonal, which is hardly expected in a realistic 
nuclear system. 
We also show that the approximation is the same order 
as that of truncation of the expansion usually done in the 
Hermitian type boson expansion theory.
%
\newpage
\section{Introduction}
\hspace*{\parindent}
The boson expansion theory (BET) \cite{KM}-\cite{RS}
has been widely used as a many-body technique to analyze 
the anharmonic effects in nuclear collective motions 
beyond the random phase approximation. 
This theory is a systematic method to transform the eigenvalue 
problem in the fermion space into the one in the boson space.
With this method, an arbitrary fermion pair operator is expressed 
as an expansion form of boson operators 
which enables us to calculate the wave functions more easily.

However BET is not a complete theory and contains some problems 
to be solved.
First one must apply the theory to the limited subspace of the
original fermion space which we call the truncated fermion subspace.
Otherwise, the boson space on which the transcribed 
boson operators act contains an unphysical part having no 
one-to-one correspondence to the original fermion space.
One cannot give the exact prescription how the limited fermion 
subspace should be selected in general.
Second, closely related to the above problem, one must adopt the 
so-called closed algebra approximation \cite{KM}.
The validity of this approximation is not clear, although it is 
indispensable for BET.
Third the convergence of the expansion for the non-collective
mode is not assured.
\footnote[2]
{To solve these problems we proposed an alternative mapping 
theory in which only the collective modes were mapped into 
the boson modes. However the purpose of the present paper is 
not to discuss about this issue. The reader may refer to 
Refs.\cite{TM} and \cite{TKM} if he has interest in it.}      

Concerning the last point, Dyson-type (D-type) BET \cite{Dy}-\cite{T-T} 
has the characteristic feature compared with the others such as 
Holstein-Primakoff-type (HP-type) BET \cite{HP}-\cite{SK}. 
In this theory the fermion space is mapped to the physical boson 
subspace by the special biunitary transformation so that a 
transformed pair operator has a {\it finite} expansion form of 
boson operators in compensation for the loss of Hermiticity. 
For example, the Hamiltonian in the fermion space is transformed 
by the Dyson boson mapping into the boson image of finite order, 
although the derived boson Hamiltonian is no longer Hermitian. 
Then in this theory, both the right and left eigenvalue problems
\cite{Taka}-\cite{T-T} 
must be solved to obtain the right and left eigenvectors, 
with which physical quantities like transition probabilities 
can be calculated by using Li's formula
\cite{Ha}-\cite{Lib}. 
Thus it had been thought for a long time that the calculation with 
D-type BET requires double efforts compared with that of HP-type BET. 

In order to remove this difficulty, the Hermitian treatment of 
D-type BET was proposed by Takada \cite{HERMITEa},\cite{HERMITEb}.
He claimed that the Hermitian matrix elements of the HP-type boson 
Hamiltonian can be calculated from the matrix elements of the 
non-Hermitian boson Hamiltonian of D-type BET by means of the so called 
Hermitization formula. 
If this formula works well, we no longer need to solve the right 
and left eigenvalue problems so that the efforts of the numerical 
calculation is greatly reduced and besides we can take the advantage 
of the finiteness of expansion. 
With the help of the Hermitian treatment, D-type BET has been 
applied to many nuclei in the various range of the nuclear chart
\cite{TTT}-\cite{TSb}. 

However the Hermitian treatment does not always hold rigorously. 
It has been remarked that it is not generally an exact method but 
is considered to be valid in good approximation when the truncated
fermion space is mapped to the physical boson subspace.
\cite{TSa},\cite{TSb}. 
Actually the Hermitian treatment was compared with the exact 
treatment of D-type BET through the numerical calculations 
in some realistic cases and the fairly good agreement was obtained.  
It was therefore concluded that the Hermitian treatment would be 
reliable enough in realistic calculations \cite{TSa},\cite{TSb}. 
However it is not yet clarified theoretically to what extent
the approximation is good in general case.

In the present paper we investigate the validity of the Hermitian 
treatment of the D-type BET from the theoretical viewpoint.
In \S2 we present a brief review of the Hermitian treatment of
D-type BET.
Here we first give the essence of BET necessary for later 
discussions and next we recapitulate the prescription of 
the Hermitian treatment.
In \S3, starting with the exact relation between the mapped 
boson operator in D-type BET and that in the HP-type BET, 
we discuss the validity of the basic assumption from which 
the Hermitization formula is derived. 
We show that the assumption does not hold except for the special 
case where the norm matrix of the multi-phonon states is diagonal.
We also show the degree of the approximation 
in the Hermitization formula in general cases explicitly.
Finally the summary of this paper is given in \S4.
%
%
\section{Brief review of the Hermitian treatment of Dyson boson
         expansion theory}
\subsection{Boson mapping theory}
\hspace*{\parindent}
In the realistic BET, the multi-phonon subspace 
$\{\to i \ket \}$ of the fermion space is mapped onto 
the ideal boson space $\{ \to i ) \}$,
where $\to i \ket$ is a multi-phonon state representing 
the multi-excitations of correlated pair modes and  
$\to i )$ is a corresponding ideal boson state \cite{KM}.
\footnote[2]
{In general the orthonormalized basis vector $\to i )$ is a 
product state of the relevant boson operators\cite{KT} 
or its linear combination using the Clebsch-Gordan coefficients, 
the boson CFP and so on \cite{Taka}. 
For simplicity we call the state $\to i)$ {\it the ideal boson state} 
in later discussions.}

It is well known that ideal boson states are orthonormalized as 
$(i \to j)=\delta_{ij}$, while multi-phonon states are not generally.
To fulfil the one-to-one correspondence between the multi-phonon space
and the ideal boson space, we first prepare the orthonormalized fermion 
basis vector 
\begin{equation}
 \label{eqn:onfbs}
 \to a \kket = n_a^{-1/2} \sum_i u_a^i \to i \ket \con
\end{equation}
by solving the eigenvalue equation for the norm matrix $Z^2$
\begin{equation}
 \label{eqn:norm}
 \sum_j (Z^2)_{ij}u_a^j=n_a u_a^i \con 
  (Z^2)_{ij} = \bra i \to j \ket \con 
\end{equation}
where $n_a$ and $u_a^i$ are the eigenvalue and the eigenvector 
respectively.
With the definition of the corresponding orthonormalized boson 
basis vector 
\begin{equation}
 \label{eqn:onbbs}
 \to a )\!) = \sum_i u_a^i \to i)  \con
\end{equation}
the HP-type mapping operator
\begin{equation}
 \label{eqn:U}
 U=\sum_{a\neq a_0} \to a )\!) \bbra a \to
\end{equation}
and the biunitary D-type mapping operator
{
\setcounter{enumi}{\value{equation}}
\addtocounter{enumi}{1}
\setcounter{equation}{0}
\renewcommand{\theequation}{\arabic{section}.\theenumi\alph{equation}}
\begin{equation}
 \label{eqn:U1}
 U_1=\sum_{a\neq a_0}n_a^{1/2} \to a )\!) \bbra a \to
\end{equation}
\begin{equation}
 \label{eqn:U2}
 U_2=\sum_{a\neq a_0}n_a^{-1/2} \to a )\!) \bbra a \to
\end{equation}
\setcounter{equation}{\value{enumi}}
}
\hspace{-5mm}
are introduced.
It is noticed that the states with zero eigenvalue($n_{a_0}=0$)
are excluded in Eqs.(\ref{eqn:U})-(\ref{eqn:U2}).
The state $ \to a_0 )\!) $ is called the unphysical state
because it has no correspondence to the original fermion state. 
Generally speaking, the ideal boson states $\ \to i) $ is not always 
physical. 
However when we treat the well-limited fermion subspace, we will have 
no zero eigenvalue of the norm matrix and $\to i) $ can be regarded 
as physical \cite{Taka},\cite{KT}.
Hereafter we assume the ideal boson state is physical to avoid 
the unnecessary ambiguities and complexity in later discussions. 

With the mapping operators defined above, an arbitrary fermion operator 
$O_F$ is transformed into the HP-type boson image $O_{HP}$ and 
the D-type one $O_D$ as
\begin{equation}
 \label{eqn:Oimage}
 O_{HP}=UO_FU^\dagger \con O_D=U_1O_FU_2^\dagger \p
\end{equation}
If we regard the operator $O_F$ as the fermion Hamiltonian $H_F$,
we are able to obtain the HP-type boson Hamiltonian $H_{HP}$ 
and the D-type boson Hamiltonian $H_D$
with which we can derive the eigenvalue problems in the boson space,
{
\setcounter{enumi}{\value{equation}}
\addtocounter{enumi}{1}
\setcounter{equation}{0}
\renewcommand{\theequation}{\arabic{section}.\theenumi\alph{equation}}
\begin{eqnarray}
 \label{eqn:HPwf}
 \left(H_{HP}-E_\lambda \right) \to \Psi_\lambda ) &=& 0  \con
 \\
 \label{eqn:Drwf}
 \left(H_{D}-E_\lambda \right) \to \psi_\lambda ) &=& 0   \con
 \\
 \label{eqn:Dlwf} 
 ( \phi_\lambda \to \left(H_{D}-E_\lambda \right) &=& 0   \p
\end{eqnarray}
\setcounter{equation}{\value{enumi}}
}
\hspace{-4.5mm}
We should notice that in the D-type theory both the right and left 
eigenvalue problems of Eqs.(\ref{eqn:Drwf}) and (\ref{eqn:Dlwf}) 
must be solved because of the non-Hermiticity of $H_D$. 

In the actual numerical calculations, we adopt one set of 
boson basis vectors to obtain a concrete representation of the 
eigenvalue equations.
Although any choice may be allowed for the representation basis set 
so long as they are orthonormalized, we ordinary take the set of 
ideal boson states $\{ \to i) \}$ which correspond to the 
multi-phonon states $\{ \to i \ket \}$.
Then we get the matrix representation of Eqs.(\ref{eqn:HPwf}),
(\ref{eqn:Drwf}) and (\ref{eqn:Dlwf}) as
{
\setcounter{enumi}{\value{equation}}
\addtocounter{enumi}{1}
\setcounter{equation}{0}
\renewcommand{\theequation}{\arabic{section}.\theenumi\alph{equation}}
\begin{eqnarray}
 \sum_j (h_{ij}^{HP}-E_\lambda \delta_{ij}) \alpha_j^{(\lambda)} &=& 0 \con
 \label{eqn:HPmt}
 \\
 \sum_j (h_{ij}^{D}-E_\lambda \delta_{ij}) \beta_j^{(\lambda)} &=& 0  \con
 \label{eqn:Drmt}
 \\
 \sum_i  \gamma_i^{(\lambda)*}(h_{ij}^{D}-E_\lambda \delta_{ij}) &=& 0 \con
 \label{eqn:Dlmt} 
\end{eqnarray}
\setcounter{equation}{\value{enumi}}
}
\hspace{-5.5mm}
where $h_{ij}^{HP}=(i \to H_{HP} \to j)$ and 
$h_{ij}^{D}=(i \to H_{D} \to j)$, 
and the coefficients $\alpha_i^{(\lambda)}, \beta_i^{(\lambda)},
\gamma_i^{(\lambda)}$ are the representation of the 
eigenvectors in this basis as follows :
{
\setcounter{enumi}{\value{equation}}
\addtocounter{enumi}{1}
\setcounter{equation}{0}
\renewcommand{\theequation}{\arabic{section}.\theenumi\alph{equation}}
\begin{eqnarray}
 \label{eqn:HPev}
 \to \Psi_\lambda ) &=& \sum_i \alpha_i^{(\lambda)} \to i ) \con
 \\
 \label{eqn:Drev}
 \to \psi_\lambda ) &=& \sum_i \beta_i^{(\lambda)} \to i ) \con
 \\
 \label{eqn:Dlev}
 (\phi_\lambda \to &=& \sum_i \gamma_i^{(\lambda)*} (i \to \p
\end{eqnarray}
\setcounter{equation}{\value{enumi}}
}
\hspace{-5mm}
It is needless to say that the HP-type matrix $h_{ij}^{HP}$ is 
Hermitian while the D-type matrix $h_{ij}^{D}$ is not generally so.
The non-Hermiticity of the matrix $h_{ij}^{D}$ can be ascribed to 
the fact that the ideal boson state $\to i)$ is not properly-normalized 
in the sense of the D-type theory.
Here we define {\it the properly-normalized states in the D-type theory} 
as follows.

Let $\to \tilde{i} \ket$ be the fermion basis vector which is related with 
the ideal boson basis state as
\begin{equation}
\label{eqn:Fi}
\to \tilde{i} \ket = U^\dagger \to i)   \p
\end{equation}
Then we can introduce a pair of D-type state vectors $\to i)_R$ and 
${}_L(i \to$ via $\to \tilde{i} \ket$ as
\begin{equation}
 \label{eqn:Di}
  \left\{
   \begin{array}{ccc}
     \to i )_R &= U_1 \to \tilde{i} \ket &= U_1 U^\dagger \to i)    \con
     \\
     \\
     {}_L (i \to &= \bra \tilde{i} \to U_2^\dagger &= (i \to U U_2^\dagger   \p
   \end{array}
  \right. 
\end{equation}
These state vectors are not Hermitian conjugate each other i.e.
${}_L (i \to \neq \{ \to i )_R \}^\dagger$, 
but they are properly-normalized in the sense that they are obtained 
from a common fermion basis vector $\to \tilde{i} \ket$ with the D-type 
mapping operator
\cite{Taka}-\cite{T-T}, \cite{HERMITEb}.

With the use of the properly-normalized basis vector, the matrix element
of the D-type boson Hamiltonian reproduces that of the HP-type 
boson Hamiltonian:
\begin{equation}
 \label{eqn:Hmt}
 {}_L (i \to H_D \to j )_R
 =\bra \tilde{i} \to H_F \to \tilde{j} \ket
 =(i \to H_{HP} \to j).
\end{equation}
Then if the explicit expression of the properly-normalized basis vector
could be easily obtained, we can treat the Hermitian eigenvalue problem 
in the D-type theory and there is no longer need to solve the right and 
left eigenvalue problems.
This is the essential idea of the Hermitian treatment which will be
discussed in the following subsection.
%
\vskip 15mm
\subsection{Hermitian treatment of Dyson boson expansion theory} 
\vskip 2mm
\hspace*{\parindent}
In this subsection we recapitulate the contents of the Hermitian treatment
\cite{HERMITEa},\cite{HERMITEb} 
and the vital assumptions used in the proof.
First we notice that from the Hermitian conjugate relation 
in the fermion space 
\begin{equation}
 \label{eqn:Ofmt}
 \bra \tilde{i} \to O_F \to \tilde{j} \ket
 = \bra \tilde{j} \to O_F^\dagger \to \tilde{i} \ket^*
\end{equation}
the corresponding relation in the boson space
\begin{equation}
 \label{eqn:Obmt}
 {}_L(i \to O_D \to j )_R
 ={}_L(j \to \ov O_D \to i )_R^* \con  
 \ov O_D = U_1 O_F^\dagger U_2^\dagger      \con
\end{equation}
should hold.
Next we {\it assume} that the properly-normalized basis states (\ref{eqn:Di}) 
 are proportional to the ideal boson states as
\begin{equation}
\label{eqn:RLbs}
 \left\{
  \begin{array}{l}  
    \to i )_R = k_i \to i)          \con
    \\
    \\
    {}_L(i \to = k_i^{-1} (i \to   \con
  \end{array}
 \right.  
\end{equation}
where the coefficients $k_i$ is assumed to be real and positive without loss 
of generality.
The substitution of Eq.(\ref{eqn:RLbs}) to Eq.(\ref{eqn:Obmt}) soon gives
\begin{equation}
 \label{eqn:Kij2}
 \left( {k_j \over k_i} \right)^2 
 = {(j\to \ov O_D \to i )^* \over (i \to O_D \to j)}
\end{equation}
so that the main formula
\begin{equation}
 \label{eqn:ODhr}
 {}_L(i \to O_D \to j)_R
 =(i \to O_D \to j) 
 \left[ {(j \to \ov O_D \to i)^* \over (i \to O_D \to j)} \right]^{1/2}
\end{equation}
can be derived.
In the above formula the matrix element of $O_D$ between the 
properly-normalized basis states is expressed in terms of the ones 
between the ideal boson states as we have hoped.
If we take $H_D$ as $O_D$ in particular, remembering 
$\ov H_D = U_1 H_F^\dagger U_2^\dagger = U_1 H_F U_2^\dagger = H_D$,
we can obtain the so called Hermitization formula of the Dyson boson 
Hamiltonian
\begin{equation}
 \label{eqn:HDhr}
 {}_L(i \to H_D \to j)_R
 =(i \to H_D \to j) 
 \left[ {(j \to H_D \to i)^* \over (i \to H_D \to j)} \right]^{1/2} \p
\end{equation}
If this formula is the exact one, we can calculate the Hermitian matrix 
of the HP-type boson Hamiltonian from the non-Hermitian D-type Hamiltonian 
$H_D$, taking the best advantage of the finiteness of the expansion.
(See Eq.(\ref{eqn:Hmt}).)
That is the outline of the Hermitian treatment
\cite{HERMITEa},\cite{HERMITEb}.

Needless to say, the key assumption in the above derivation is 
Eq.(\ref{eqn:RLbs}).
In Ref.\cite{HERMITEb}, Takada has presented the condition under 
which Eq.(\ref{eqn:RLbs}) is satisfied and the Hermitian treatment 
is justified.
In the latter part of this subsection, after following the essential 
part in Ref.\cite{HERMITEb}, the problems contained in his derivation 
is pointed out. 

The ideal boson states $\to i)$
are usually specified by the set of quantum numbers, for example,  
the boson number, the angular momentum and its projection etc. 
In other words, the ideal boson states are simultaneous eigenstates 
of the total boson number operator $\wh N_B$ and a set of the 
other operators $C_{HP}(k) (k=1,2,\cdots,K)$ which commute 
with $\wh N_B$: 
\begin{equation}
\label{eqn:CHP}
 \left\{
  \begin{array}{l}  
   \wh N_B \to i) = {\cal N}_i \to i )     \con
    \\
    \\
   C_{HP}(k) \to i) = {\cal M}(k)_i \to i)   \con
 \end{array}
 \right.  
\end{equation}
where $K$ is the number of $C_{HP}(k)$ needed for this purpose.
It is always possible to find such operators as $C_{HP}(k)$ in various 
kinds of calculations. 
As the operator $C_{HP}(k)$ is to be the HP-type boson image of 
a certain fermion operator $C_{F}(k)$, the corresponding D-type boson 
image $C_{D}(k)$ must exist. 
Taking Eq.(\ref{eqn:Di}) into account, 
Eq.(\ref{eqn:CHP}) soon gives the right and left eigen-equations for 
the properly-normalized basis states
\begin{equation}
\label{eqn:CD}
 \left\{
  \begin{array}{l}  
    C_{D}(k) \to i)_R = {\cal M}(k)_i \to i)_R     \con
    \\
    \\
    {}_L(i \to C_{D}(k) = {}_L (i \to {\cal M}(k)_i   \p
  \end{array}
 \right.  
\end{equation}
After the above preparation, 
we adopt the following proposition stated in Ref.\cite{HERMITEb}:

\noindent
proposition [A]: 
{\it
According to the general property of the HP-type and D-type boson 
mappings}
 \cite{JDFJ}, 
{\it
the boson images of a fermion operator 
commutable with the fermion number operator are the same 
for both the types.}
\vskip 2mm
\noindent
This proposition immediately gives the relation 
\begin{equation}
 \label{eqn:CD-CHP}  
 C_D(k)=C_{HP}(k)   \con
\end{equation}
because the operator $C_{HP}(k)$ commutes with $\wh N_B$ 
so that the fermion counterpart $C_{F}(k)$ also commutes with 
the fermion number operator.
Then we can get from Eqs.(\ref{eqn:CD}) and (\ref{eqn:CD-CHP}) 
\begin{equation}
\label{eqn:CHPrl}
 \left\{
  \begin{array}{l}  
    C_{HP}(k) \to i)_R = {\cal M}(k)_i \to i)_R    \con
    \\
    \\
    {}_L(i \to C_{HP}(k) = {}_L (i \to {\cal M}(k)_i   \p
  \end{array}
 \right.  
\end{equation}
Remembering the assumption that the degeneracy of $\to i)$ is completely
resolved, the comparison of Eqs.(\ref{eqn:CHP}) and (\ref{eqn:CHPrl})
leads to the equations
\begin{equation}
\label{eqn:bsrl}
 \left\{
  \begin{array}{l}  
    \to i )_R = k_i \to i)          \con
    \\
    \\
    {}_L(i \to = k_i^{-1} (i \to   \con
  \end{array}
 \right.  
\end{equation}
with the appropriate constant $k_i$, 
which is nothing but Eq.(\ref{eqn:RLbs}). 
As we can see in the above derivation, not so strict condition 
is to be required for the use of the Hermitian treatment. 
{\it
So long as a set of ideal boson states $\{ \to i) \}$ is prepared  
as the eigenstates of the total boson number operator $\wh N_B$, 
the Hermitization formula could be used as the exact one.
}
This claim is the main conclusion of Ref.\cite{HERMITEb}.

The proposition [A] is the key point in the proof of Eq.(\ref{eqn:CD-CHP}).
It is based on the work of Janssen et al.\cite{JDFJ}, which has given 
explicitly the boson representation of the fermion pair operator 
in both the HP-type and the D-type boson mapping as follows:
%
\begin{equation}
\label{eqn:Janssen}
{\rm HP}:
 \left\{
  \begin{array}{l}
   a_\alpha^\dagger a_\beta^\dagger \longrightarrow 
   \wh P \Big( b^\dagger \cdot \sqrt{1-(b^\dagger b)}\ 
   \Big)_{\alpha \beta} \wh P
  \\
   a_\beta a_\alpha \longrightarrow 
   \wh P \Big( \sqrt{1-(b^\dagger b)}\cdot b \Big)_{\alpha \beta} \wh P
  \\
   a_\alpha^\dagger a_\beta \longrightarrow 
   \wh P (b^\dagger b)_{\beta \alpha} \wh P
  \end{array}
 \right.
,
{\rm D}:
 \left\{
  \begin{array}{l}
   a_\alpha^\dagger a_\beta^\dagger 
   \longrightarrow 
   \wh P \Big(b^\dagger-b^\dagger(b^\dagger b)
   \Big)_{\alpha \beta} \wh P
  \\
   a_\beta a_\alpha 
   \longrightarrow 
   \wh P b_{\alpha \beta} \wh P
  \\
   a_\alpha^\dagger a_\beta 
   \longrightarrow 
   \wh P (b^\dagger b)_{\beta \alpha} \wh P
  \end{array}
 \right.
.
\end{equation}
Here $ \wh P$ denotes the projection operator to the physical
boson subspace.%
(For the meaning of the other symbols 
and the more accurate expression, see Ref.\cite{JDFJ}.)

It is  supposed that the proposition [A] would be derived in 
Ref.\cite{HERMITEb} from the fact that the boson image of
the $(a^\dagger a)$-type operator has the same form 
both in the HP-type and the D-type theories in 
Eq.(\ref{eqn:Janssen}).
However this fact is insufficient to justify the 
proposition [A] as is shown below.

First we should notice that Eq.(\ref{eqn:Janssen}) has been 
derived in the very case when {\it the whole fermion space} 
is mapped onto the antisymmetrized physical boson subspace 
in the particle pair representation.
In this case the ideal boson states generally contain the 
unphysical component and hence the projection operator $\wh P$ 
cannot be considered as unity. 
It is expressed by the very complicated  and non-convergent 
form of the boson operators, which makes BET itself unworkable
and also spoil the finite form of D-type BET.
Therefore in a realistic BET, we must truncate the fermion space 
beforehand so that the projection operator $\wh P$ can be regarded 
as unity, as was mentioned in \S 1.
We cannot apply Eq.(\ref{eqn:Janssen}) to the realistic BET
since the mapping operator is quite different from that used 
by Janssen in this case.

Second the operator commutable with the fermion number operator 
does not limited to the $ (a^\dagger a)$-type of pair operator. 
If $C_F(k)$ contains the operator of the form 
$ (a^\dagger a^\dagger) (a a) $,
for example, its boson image of the HP-type and that of the D-type
are not identical with each other in the realistic BET.

In the following section, we will investigate the key identity 
(\ref{eqn:CD-CHP}) on the firmer ground, and show that 
the key assumption (\ref{eqn:RLbs}) does not hold generally. 
%
\setcounter{equation}{0}
\section{Hermitian treatment as approximate technique} 
\subsection{Rigorous treatment}
\hspace*{\parindent}
In this subsection, starting from the exact relation between the HP-type 
boson image and the D-type one, we re-examine the condition which allows us 
to use the Hermitian treatment as a rigorous method.

Using the definition of the boson image (\ref{eqn:Oimage}), we get 
the exact relation between two types of image, $O_{HP}$ and $O_{D}$, 
for any fermion operator $O_F$:

\begin{equation}
 \label{eqn:ODexact}
 O_D = \wh Z_B O_{HP} \wh Z_B^{-1} 
\end{equation}
with the definition  
\begin{equation}
 \label{eqn:ZBdef}
 \wh Z_B = U_1 U^\dagger = ( U U_2^\dagger )^{-1} . 
\end{equation}
By using Eqs.(\ref{eqn:onfbs})-(\ref{eqn:U1}), the boson operator 
$\wh Z_B$ can be expressed as 
\begin{eqnarray}
 \label{eqn:ZBmt}
 \wh Z_B 
         &=& \sum_{a\neq a_0} n_a^{1/2} \to a )\!)(\!(a \to  \nonumber
 \\
         &=& \sum_{ij} Z_{ij} \to i )( j \to ,
 \end{eqnarray}
where $Z_{ij}$ is a matrix element of the square root of the multi-phonon
norm matrix $Z^2$. 

In order that the key identity (\ref{eqn:CD-CHP}) may hold, 
the condition
\begin{equation}
\label{eqn:CHP-ZB}
[C_{HP}(k),\wh Z_B]=0
\end{equation}
is clearly necessary and sufficient.
Since a set of the operators $\{ C_{HP}(k);k=1,2,\cdots,K \}$ labels 
the state $\to i )$ with a fixed boson number completely, 
this equation means that $\wh Z_B$ is diagonal, that is 
\begin{equation}
 \label{eqn:Znd=0}
 Z_{ij}=0 \qquad  {\rm for\ any\ } i\neq j    \p
\end{equation}
Thus the key identity (\ref{eqn:CD-CHP}) becomes invalid if the norm
matrix of the multi-phonon states has non-diagonal matrix elements.
In the special system which has an orthogonal set of multi-phonon states 
(for example, the SU(3) model \cite{Takb},\cite{LKD}, the SU(4) model 
\cite{IT} and so on), the key identity (\ref{eqn:CD-CHP}) is satisfied 
so that the Hermitian treatment is guaranteed to be exact.
Generally speaking, however, multi-phonon states are not mutually 
orthogonal due to the Pauli effect. 
Hence the identity (\ref{eqn:CD-CHP}) does not hold,
which accordingly leads to the conclusion that the key assumption
(\ref{eqn:RLbs}) is not correct generally.  

The above result can be derived more directly from the relation
between the D-type basis vector and the HP-type one.
By using Eqs.(\ref{eqn:Di}),(\ref{eqn:ZBdef}) and (\ref{eqn:ZBmt}), 
we get
\begin{equation}
 \label{eqn:D-HPbs}
 \left\{
  \begin{array}{ccc}  
    \to i )_R &= \wh Z_B \to i) &= {\displaystyle \sum_j Z_{ij}} \to j)    \con
    \\
    {}_L(i \to &= (i \to \wh Z_B^{-1} &= {\displaystyle \sum_j Z_{ij}^{-1}}
     (j \to   \con
  \end{array}
 \right.  
\end{equation}
which clearly show that the key assumption of the Hermitian treatment
(\ref{eqn:RLbs}) holds only when the norm matrix is diagonal.

In order for the Hermitian treatment to be justified, 
there is no need to require the identity (\ref{eqn:CD-CHP}).
A brief look at the contents of the subsection 2.2 makes us understand 
that the foundation of the Hermitization formula is Eq.(\ref{eqn:RLbs})
i.e. the proportionality of the ideal boson state to 
the properly-normalized D-type basis state.
Here we should notice that the representation basis set
is not necessarily restricted to that of 
ideal boson states $\{ \to i) \}$. 
If we find an appropriate set of basis states 
$\{\to \alpha) \}$ which satisfies the proportionality
\begin{equation}
 \label{eqn:alpha}
 \left\{
  \begin{array}{cc}
   \to \alpha )_R & = k_\alpha \to \alpha)            \con
   \\
   {}_L(\alpha \to & = k_\alpha^{-1} ( \alpha \to     \con
  \end{array} 
 \right.
\end{equation}
we are able to use the Hermitization formula with regard to this basis.

Then how can we obtain such a kind of basis set ?
Since the new basis state $\to \alpha)$ is given by a linear
combination of the basis state $\to i)$,  Eq.(\ref{eqn:D-HPbs})
also holds for $\to \alpha)$:
\begin{equation}
 \label{eqn:alphaD}
 \left\{
  \begin{array}{cc}
   \to \alpha)_R &= \ \ \wh Z_B \to \alpha )              \con
  \\
  {}_L(\alpha \to &= (\alpha \to \wh Z_B^{-1}        \p
  \end{array}
 \right.
\end{equation}
Combination of this equation and Eq.(\ref{eqn:alpha}) directly gives
\begin{equation}
 \label{eqn:esZ_B}
 \left\{
  \begin{array}{cc}
   \wh Z_B \to \alpha ) &= k_\alpha \to \alpha)    \con
   \\
   (\alpha \to \wh Z_B^{-1} &= k_\alpha^{-1} (\alpha \to    \con
  \end{array}
 \right.
\end{equation}
which means that the expected basis state $\to \alpha)$ 
must be an eigenstate of the operator $\wh Z_B$. 
The new basis state $\to \alpha)$ is nothing but the basis 
state $\to a )\!)$ for the physical boson subspace defined by 
Eq.(\ref{eqn:onbbs}):
\begin{equation}
 \label{eqn:s-alpha}
 \to \alpha)= \to a )\!) = \sum_i u_a^i \to i )    \con
\end{equation}
where $\{u_a^i\}$ is a unitary matrix to diagonalize the norm
matrix $Z^2$. (See Eq.(\ref{eqn:norm}).) 
Therefore it is possible to prepare a set of basis states 
which enables us to use the Hermitian treatment as an exact
method, if and only if we select the representation making 
the norm matrix $Z^2$ diagonal. 
However we usually never do such a thing because the efforts needed 
to solve the eigenvalue equation of the norm matrix completely lose 
the merit of performing the boson mapping.
%
\subsection{Approximation in the conventional treatment}
%
\hspace*{\parindent}
In the previous subsection we have shown that the basic assumption 
of the Hermitian treatment of D-type BET does not hold generally. 
Then how can we explain the fact that it is a good 
approximation to the exact treatment in some realistic cases 
as was shown numerically in Refs. \cite{TSa} and \cite{TSb}?
In this subsection, 
we show that the result of the Hermitian treatment is approximately 
valid in spite of the breakdown of the basic assumption.

First we express the norm matrix element as
\begin{equation}
 \label{eqn:Z2}
 (Z^2)_{ij}=\delta_{ij}-Y_{ij}   \con
\end{equation}
where the matrix element $Y_{ij}$ represents the non-orthonormality 
of the multi-phonon states 
and stems from the Pauli effect between fermions belonging 
to the different phonon operators\cite{KT},\cite{SK}.
If we truncate the multi-phonon subspace to some extent, 
the absolute value of $Y_{ij}$ becomes so small that 
it can play a role of a small parameter\cite{KT},\cite{SK}. 
Next we introduce the norm operator $\wh Z_B^2$ 
in the boson space 
\begin{equation}
 \label{eqn:ZB2}
 \wh Z_B^2 = (\wh Z_B)^2 = 1_B - \wh Y_B
\end{equation}
with
\begin{equation}
 \label{eqn:1B}
 1_B=\sum_i \to i)(i \to  \con
 \wh Y_B=\sum_{ij}Y_{ij} \to i)(j \to    \p
\end{equation}
In the following we expand the operators $\wh Z_B$ 
and $\wh Z_B^{-1}$ with regard to the operator $\wh Y_B$ and 
write them up to $O(\wh Y_B)$: 
\begin{equation}
 \label{eqn:ZB1}
 \wh Z_B^{\pm 1} = (1_B-\wh Y_B)^{\pm 1/2}
 =1_B \mp {1\over 2} \wh Y_B +O(\wh Y_B^2)     \p
\end{equation}
Then the relation between the D-type and HP-type boson images of 
the Hamiltonian can be expressed as
\begin{equation}
 \label{eqn:HD}
 H_D=\wh Z_B H_{HP} \wh Z_B^{-1}
 =H_{HP} -{1\over 2} [\wh Y_B, H_{HP}] + O(\wh Y_B^2)     \con
\end{equation}
whose matrix element between 
the ideal boson basis states $\to i)$ and $\to j)$ becomes
\begin{equation}
 \label{eqn:HDmt}
 (i \to H_D \to j)
 =
 (i \to H_{HP} \to j)-{1\over 2}{ (i \to [\wh Y_B, H_{HP}] \to j)}
 +O(\wh Y_B^2)
 \p
\end{equation}
From Eq.(\ref{eqn:HDmt}) we can easily verify the expression
\begin{eqnarray}
 \label{eqn:rtHDmt}
  {(j \to H_D \to i)^* \over (i \to H_D \to j)}
   &=& 
   {(j \to H_{HP} \to i)^* -{1\over 2}(j \to [\wh Y_B, H_{HP}] \to i)^*
   \over
   (i \to H_{HP} \to j) -{1\over 2}(i \to [\wh Y_B, H_{HP}] \to j)}
   +O(\wh Y_B^2) \nonumber
 \\     \nonumber
 \\
   &=&
   1+{ (i \to [\wh Y_B, H_{HP}] \to j) \over (i \to H_{HP} \to j)}
   +O(\wh Y_B^2)
\end{eqnarray}
and the square root of it
\begin{equation}
 \label{eqn:rtHDmt2}
 \left[ {(j \to H_D \to i)^* \over (i \to H_D \to j)} \right]^{1/2}
 =
 1+{1\over 2}{ (i \to [\wh Y_B, H_{HP}] \to j) \over (i \to H_{HP} \to
 j)}
 +O(\wh Y_B^2)
 \p
\end{equation}
Multiplying Eqs.(\ref{eqn:HDmt}) and (\ref{eqn:rtHDmt2}) side by side, 
the terms of $O(\wh Y_B)$ are neatly canceled out and we get 
\begin{equation}
 \label{eqn:HDmtHP}
 (i \to H_D \to j)\left[ {(j \to H_D \to i)^* \over (i \to H_D \to j)} 
 \right]^{1/2}
 =(i \to H_{HP} \to j) +O(\wh Y_B^2) .
\end{equation}
Noticing the identity $(i \to H_{HP} \to j)={}_L(i \to H_{D} \to j)_R$, 
Eq.(\ref{eqn:HDmtHP}) makes us understand 
that the Hermitization formula (\ref{eqn:HDhr}) is justified 
under the approximation to ignore terms of $O(\wh Y_B^2)$ in general
case. 

As is well known, 
in the HP-type theory\cite{HP}-\cite{SK} 
the phonon operator (the correlated pair mode) 
is transformed to an infinite expansion form of boson operators, 
which is symbolically expressed as
\begin{equation}
 \label{eqn:bHP}
 b^\dagger + c_1 Y b^\dagger b^\dagger b 
 + c_2 Y^2 b^\dagger b^\dagger b^\dagger b b + \cdots    \con
\end{equation}
where $b^\dagger$ ($b$) is a boson creation (annihilation) operator, 
$c_k(k=1,2,...)$ is a numerical factor, 
and $Y$ represents a small parameter of $O(\wh Y_B)$.
Therefore except the special case given in the subsection 3.1, 
the application of the Hermitization formula in the D-type theory 
is nearly equivalent to the approximation which truncates the infinite 
expansion to the finite order up to $O(Y)$ in the HP-type theory. 

Now we can answer the question stated at the top of this subsection.
In the realistic systems treated in Refs. \cite{TSa} and \cite{TSb}, 
the above $Y$-parameter is probably small enough to neglect 
the higher order terms in r.h.s. of Eq.(\ref{eqn:HDmtHP}).
However it should be noted that the HP-type calculation up to $O(Y)$ 
will also give nearly the same results in this case.
%
%
\section{Summary}
\hspace*{\parindent}
In the present paper, we have investigated the validity of the 
Hermitian treatment in D-type BET from the theoretical point of view. 
To make the discussion clear we confined ourselves to the case in 
which the ideal boson states have no unphysical component.
From the exact relation between the D-type theory and 
the HP-type one, we have shown the following:
{\it the key assumption (\ref{eqn:RLbs}) of the Hermitian treatment 
holds only if the norm matrix $Z^2$ is diagonal.}
Hence the statement in Ref.\cite{HERMITEb} that the key assumption 
(\ref{eqn:RLbs}) holds for a boson state with a 
fixed total boson number should be replaced by the above statement.

In a realistic case the norm matrix of multi-phonon states never 
becomes diagonal.
The situation is unchanged even if we estimate the norm matrix
by using the so called closed algebra approximation.
One might therefore suppose that the validity of the Hermitian treatment
is very doubtful. 
Nevertheless, the result of the treatment i.e. the Hermitization formula 
holds approximately. 
We have shown it by expanding the norm matrix with the parameter exhibiting 
the Pauli effect on multi-phonon states. 
It happens that the first order terms with respect to this parameter 
are canceled out in the Hermitization formula. 
It is owing to this cancellation mechanism ( not to the basic assumption 
(\ref{eqn:RLbs})) that the formula is approximately valid. 
The approximation neglecting the second order terms is  
the same as that adopted in the conventional HP-type BET,
although the latter method can be developed to higher order in principle.

In summary, the Hermitian treatment is no longer an exact method 
but an approximate technique which is nearly equivalent to the 
finite-truncated HP-type theory. 
As a result, the use of the Hermitization formula in the D-type 
calculation reduces the great merit of the finiteness of the expansion.
Hence we conclude that the right and left eigenvalue problems should be 
solved if we want to take the best advantage of D-type BET.
%
\clearpage


\begin{thebibliography}{99}
%
\bibitem{KM} A.Klein and E.R.Marshalek, Rev.Mod.Phys.63(1991),375
\bibitem{Ma} E.R.Marshalek, Nucl.Phys.A347(1980),253
\bibitem{RS} P.Ring and P.Schuck,
              The Nuclear Many-Body Problem (Springer,1980)
%
\bibitem{TM} K.Taniguchi and Y.Miyanishi, Prog.Theor.Phys.86(1991),151
\bibitem{TKM} K.Taniguchi, A.Kajiyama and Y.Miyanishi, 
              Prog.Theor.Phys.92(1994),975
%
\bibitem{Dy} F.J.Dyson, Phys.Rev.102(1956)1217,1230
\bibitem{JDFJ} D.Janssen, F.D\"onau, S.Frauendorf and R.V.Jolos,
               Nucl.Phys.A172(1971),145
%
\bibitem{Tam}  T.Tamura, Phys.Rev.C28(1983),2480
\bibitem{Taka} K.Takada, Nucl.Phys.A439(1985),489
\bibitem{T-T-T} K.Takada, T.Tamura and S.Tazaki, Phys.Rev.C31(1985),1948
\bibitem{T-T} K.Takada and S.Tazaki, Nucl.Phys.A448(1986),56
%
\bibitem{HP} T.Holstein and H.Primakoff, Phys.Rev.58(1940),1098
\bibitem{MYT} T.Marumori, M.Yamamura and A.Tokunaga,
              Prog.Theor.Phys.31(1964),1009
\bibitem{LH} S.G.Lie and G.Holzwarth, Phys.Rev.C12(1975),1035
\bibitem{MTS} T.Marumori, K.Takada and F.Sakata,
              Prog.Theor.Phys.Supplement,71(1981),1
%
\bibitem{KT} T.Kishimoto and T.Tamura, Phys.Rev.C27(1983),341
\bibitem{SK} H.Sakamoto and T.Kishimoto, Nucl.Phys.A486(1988),1
%
\bibitem{Ha} F.J.W.Hahne, Phys.Rev.C23(1981),2305
\bibitem{Lia} C.T.Li, Phys.Lett.120B(1983),251
\bibitem{Lib} C.T.Li, Nucl.Phys.A417(1984),37
%
\bibitem{HERMITEa} K.Takada, Phys.Rev.C34(1986),750
\bibitem{HERMITEb} K.Takada, Phys.Rev.C38(1988),2450
%
\bibitem{TTT} H.Tsukuma, H.Thorn and K.Takada, Nucl.Phys.A466(1987),70
\bibitem{TY} K.Takada and K.Yamada, Nucl.Phys.A462(1987),561
\bibitem{YTa} K.Yamada and K.Takada, Nucl.Phys.A480(1988),143
\bibitem{TYT} K.Takada, K.Yamada and H.Tsukuma, Nucl.Phys.A496(1989),224
\bibitem{YTT} K.Yamada, K.Takada and H.Tsukuma, Nucl.Phys.A496(1989),239
\bibitem{YTb} K.Yamada and K.Takada, Nucl.Phys.A503(1989),53
%
\bibitem{TSa} K.Takada and Y.R.Shimizu, Nucl.Phys.A523(1991),354
\bibitem{TSb} K.Takada and Y.R.Shimizu, Prog.Theor.Phys.95(1996),1121
%
\bibitem{Takb} K.Takada, Nucl.Phys.A490(1988),262
\bibitem{LKD} S.Y.Li, A.Klein and R.M.Dreizler, J.Math.Phys.11(1970),975 
\bibitem{IT} S.Ichikawa and K.Takada, Nucl.Phys.A552(1993),193
%
\end{thebibliography}
\end{document}